\documentclass[aps,prb,twocolumn,preprintnumbers]{revtex4}

\usepackage{amsmath}
\usepackage{graphicx}
\usepackage{soul}
\usepackage{amsfonts}
\usepackage{amssymb}
\usepackage{dcolumn}
\usepackage{adjustbox}
\usepackage{color}

\usepackage{tabulary,graphicx,amsmath,amsfonts,amssymb}
\usepackage[utf8x]{inputenc}
\usepackage[T1]{fontenc}
\let\citep\cite
\let\citet\cite
\usepackage{hyperref}

\usepackage{float}

\makeatletter

\begin{document}

\title{Tunable Resonant Metasurfaces Empowered by Atomically Thin Semiconductors}

\author{Alexey Ustinov$^{1,2,*}$, Ángela Barreda$^3$, Duk-Yong Choi$^4$, Tobias Bucher$^{1,2}$, Giancarlo Soavi$^1$, Thomas Pertsch$^{2,5}$ \& Isabelle Staude$^{1,2}$}

\affiliation{$^1$Institute of Solid State Physics, Friedrich Schiller University Jena, Max-Wien-Platz 1, 07743 Jena, Germany\\
$^2$Institute of Applied Physics, Abbe Center of Photonics, Friedrich Schiller University Jena, Albert-Einstein-Str. 15, 07745 Jena, Germany\\
$^3$Department of Electronic Engineering, University Carlos III of Madrid, Avda. de la Universidad, 30, 28911 Leganés, Spain\\
$^4$Laser Physics Centre, Australian National University, Canberra, ACT 2601, Australia\\
$^5$Fraunhofer Institute for Applied Optics and Precision Engineering, Albert-Einstein-Str. 7, 07745 Jena, Germany\\
~~*Corresponding author: \href{mailto:alexey.ustinov@uni-jena.de}{alexey.ustinov@uni-jena.de}~~}

\begin{abstract}
Nanophotonics has recently gained new momentum with the emergence of a novel class of nanophotonic systems consisting of resonant dielectric nanostructures integrated with single or few layers of transition metal dichalcogenides (2D-TMDs). Thinned to the single layer phase, 2D-TMDs are unique solid-state systems with excitonic states able to persist at room temperature and demonstrate notable tunability of their energies in the optical range. Based on these properties, they offer important opportunities for hybrid nanophotonic systems where a nanophotonic structure serves to enhance the light-matter interaction in the 2D-TMDs, while the 2D-TMDs can provide various active functionalities, thereby dramatically enhancing the scope of nanophotonic structures.
\par
In this work, we combine 2D-TMD materials with resonant photonic nanostructures, namely, metasurfaces composed of high-index dielectric nanoparticles. The dependence of the excitonic states on charge carrier density in 2D-TMDs leads to an amplitude modulation of the corresponding optical transitions upon changes of the Fermi level, and thereby to changes of the coupling strength between the 2D-TMDs and resonant modes of the photonic nanostructure. We experimentally implement such a hybrid nanophotonic system and demonstrate voltage tuning of its reflectance as well as its different polarization-dependent behavior.
\par
Our results show that hybridization with 2D-TMDs can serve to render resonant photonic nanostructures tunable and time-variant -- important properties for practical applications in optical analog computers and neuromorphic circuits.
\end{abstract}

\maketitle

\section{Introduction}   
For decades, TMDs have been employed as grease material in technical applications due to their low friction, as a result of the weak Van der Waals forces between their individual layers. However, when reduced to their monolayer form, TMDs exhibit a range of remarkable optical, mechanical, and electronic properties that position them as promising materials for future technologies \cite{Elshaari:20, Bogaerts:20}. Their intrinsic two-dimensional confinement and reduced dielectric screening result in a substantial exciton binding energy up to 0.5 eV, which make them stable even at room temperature \cite{Wang:15, Cheiwchanchamnangij:12, Chernikov:14}, while their direct bandgap enables intense photoluminescence (PL) \cite{Wang:12}. Additionally, TMD monolayers demonstrate strong and tunable non-linear optical effects, which allow the realization of advanced devices for all-optical modulation of polarization and wavefront shaping of light \cite{Klimmer:21, Sinelnik:24}.
\par
The unique electronic structure of TMDs allows for precise manipulation of their optical properties by externally controlling the former. In particular, their optical characteristics are highly tunable through external stimuli such as chemical doping \cite{Ryder:16}, electrical gating \cite{Seyler:15, Yu:17, Chernikov:15}, optical fields \cite{Cunningham:17, Fan:17}, mechanical strain \cite{Peng:20}, changes in dielectric environment \cite{Raja:17, Stier:16}, substrate effects \cite{Kobayashi:15}, and temperature variations \cite{Lanzillo:13, Thripuranthaka:14}, among others. This adaptability underscores the potential of monolayer TMDs in advancing next-generation photonic, electronic, and optoelectronic systems. 
\par
However, the single-atom thickness of TMD monolayers limits the light-matter interactions with the incident electromagnetic radiation. This limitation poses challenges in observing key optical phenomena, such as absorption and reflection, as well as in achieving their active response to external stimuli. Under standard conditions, their relative change under external modulation stimuli is relatively low and often requires cryogenic temperatures for these effects to become noticeable \cite{Li:21}. A way to overcome this challenge is to use hybrid systems, which rely on the coupling of TMD monolayers to nanostructures able to confine optical fields in sub-wavelength volumes of the monolayer region \cite{Chen:17, Chen:20, Loechner:21}.
\par
In this context, optical bound states in the continuum (BICs) supported by all-dielectric nanostructures present a particularly promising platform. BICs are unique modes that remain localized despite existing within the continuum spectrum of radiating states \cite{Hsu:13}. Initially introduced in quantum mechanics, their relevance has expanded to photonics, where they play a significant role in fields such as lasing, non-linear optics, sensing, chirality, and PL \cite{Kodigala:17, Bernhardt:20, Kim:22_3, Gorkunov:20, Zhu:20}. Theoretically, classical BICs are considered dark modes that neither couple to incident radiation nor emit it, exhibiting an infinite $Q$-factor \cite{Hsu:13}.
BICs can be categorized based on the mechanisms that suppress their interaction with incident radiation. Symmetry-protected BICs occur when the symmetry of the fields in BIC modes is incompatible with that of the radiating waves, preventing coupling between them \cite{Hsu:19, Koshelev:19}. In contrast, accidental BICs emerge when the continuous tuning of system parameters leads to the accidental vanishing of the coupling coefficient to radiative waves \cite{Han:19}.
\par
However, these ideal non-radiative modes are purely mathematical constructs and cannot be observed directly in experiments. In practical implementations, the BIC transforms into a leaky mode, commonly referred to as a quasi-BIC (q-BIC) state \cite{Abujetas:19}, showing high yet finite Q-factors.
\par
The coupling of TMDs with all-dielectric metasurfaces supporting q-BIC resonances at an excitonic wavelength significantly enhances the light-matter interactions within the former. Among other applications such as enhancement of PL emission, second-harmonic generation, low threshold \cite{Wu:15} and room-temperature lasing \cite{Li:17}, enhanced tuning effects are expected in hybrid nanostructures. The control of optical properties in such hybrid structures relies on dynamically altering the effective refractive index of a TMD monolayer, which in turn induces changes in its spectral response. Specifically, excitonic states in atomically thin TMDs are highly dependent on the Fermi level \cite{Li:21}, which can be manipulated by the application of an external stimulus, such as a constant electric field. At the charge-neutral point, where the densities of electrons and holes are balanced, the optical properties of a TMD monolayer are primarily governed by neutral exciton transitions. These transitions dominate because the absence of excess carriers minimizes Coulomb screening and scattering effects, allowing the exciton formation probability to remain high \cite{Li:23}, while shifted Fermi-levels would allow for these phenomena to compete with the (neutral) excitonic transitions.
\par
When an external stimulus, such as a constant electric field \cite{Zhao:20}, is applied, it alters the Fermi level of the system. This modification in the Fermi level corresponds to a change of the carrier density in a semiconductor, which in turn modifies the optical response of the monolayer through the following key mechanisms: Coulomb screening and scattering effects, charged excitons (trions) formation, Pauli blocking, bandgap renormalization, and spectral broadening \cite{Li:23}. The three latter have a weak influence on the exciton binding energy \cite{Yu:17}. Regarding the Coulomb screening and scattering effects, the additional charge carriers screen the Coulomb interaction between electrons and holes of the neutral exciton, decreasing exciton formation probability. This results in a shift of contribution in optical response from neutral to charged excitons, leading to effective spectral broadening of the excitonic line and shift of the excitonic peak. As a consequence, both emission and absorption/reflection processes are spectrally modified. The oscillator strength of trions formed with three quasi-particles, which are spectrally close to neutral excitons, is by few times of magnitude weaker at room temperature, compared to that of neutral excitons \cite{Mukherjee:15}. Trionic contribution becomes visible as a broader resonance in PL or absorption spectra only when the neutral exciton is suppressed.
\par
\begin{figure*}
\centerline{\includegraphics[width = 0.9\linewidth]{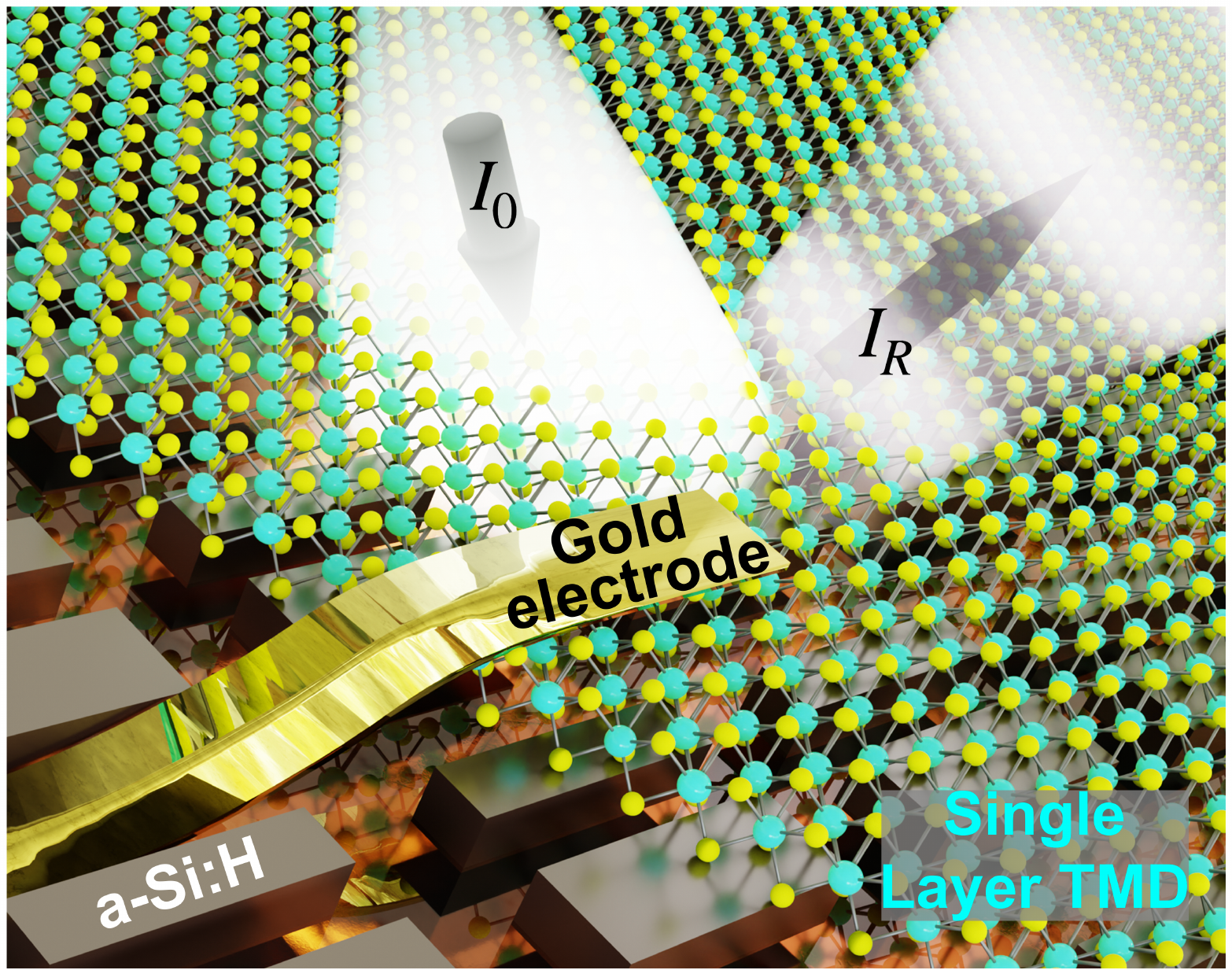}}
\caption{An artist's impression of the of hybrid resonant metasurface (not to scale). A WSe$_2$ monolayer is positioned on top of a metasurface supporting a q-BIC state within the  wavelength range of the excitonic resonance of the former. A gold electrode allows for the injection of charge carriers and thus Fermi-level control, leading to tunablility of the linear-optical properties of the hybrid metasurface.}
\label{fig1}
\end{figure*}

In this work, we demonstrate numerically and experimentally the active tuning of a hybrid system consisting of a TMD (WSe$_2$) monolayer coupled to a Mie-resonant metasurface. The conceptual appearance of the structure is represented in Fig.~${\hyperref[fig1]{\ref{fig1}}}$. Tunability is achieved through application of external electric field, enabling dynamic control of the system's linear-optical properties, specifically its absorptance and reflectance. By controlling the Fermi level in the WSe$_2$ monolayer via an external electric field, we manipulate excitonic population distribution between neutral ($A^0$) and charged ($A^{\pm}$) states. This balance between excitonic states primarily governs the complex refractive index in TMD monolayers within the visible and near-infrared ranges of wavelengths at room temperature. The enhancement of the neutral $A^0$-exciton increases the real part of the WSe$_2$ monolayer's optical conductivity, which subsequently leads to greater absorption. The coupling effect between the WSe$_2$ monolayer and the dielectric metasurface, which exhibits a q-BIC resonance at the $A^0$ exciton wavelength, leads to an enhancement of this modulation effect. As a consequence, we report a modulation depth of at least 10$\%$ in the reflection spectrum of the hybrid metasurface. The results of this work could serve as a platform for further investigation of exciton dynamics in monolayer TMDs and its interaction regimes with resonant photonic modes.

\section{Results}

\subsection{Metasurface Design and Linear Characterization}
\begin{figure*}
\centerline{\includegraphics[width = 0.9\linewidth]{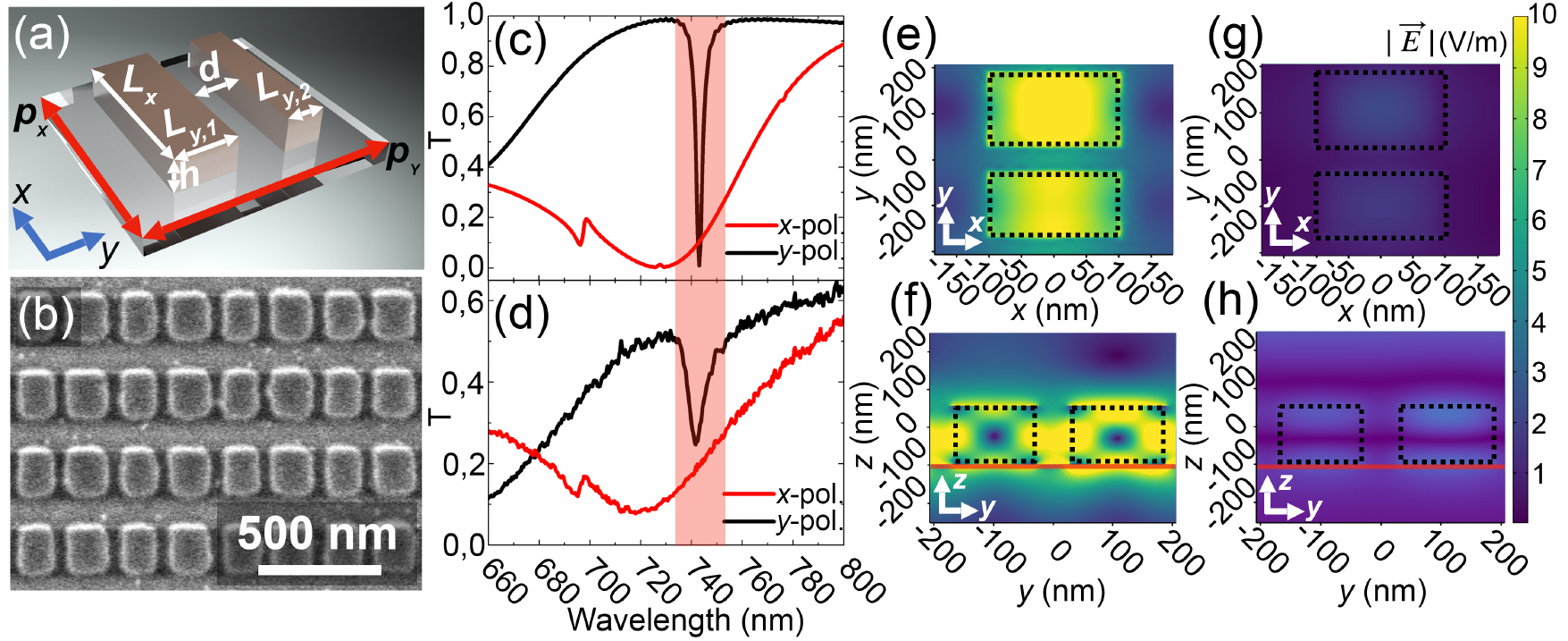}} 
\caption{The q-BIC metasurface design and characterization. (a) Sketch of the q-BIC metasurface unit cell with denoted geometrical parameters. (b) SEM image of the q-BIC metasurface area. (c) Simulated and (d) measured transmission spectra of the bare metasurface sample featuring q-BIC-resonance at 742~nm for normally incident linearly polarized light. Red and black curves are related to $x$- (off-resonant case) and $y$-polarized light (q-BIC excitation), respectively. Red shaded area outlines the excitonic band represented by neutral A$^{0}$ and charged A$^{\pm}$-trions. (e-h) Near-field maps at $\lambda = 742$~nm for the polarization of the incident radiation along the $y$- (q-BIC excitation) (e, f) and $x$-axis (off-resonant case) (g, h) in $XY$- (e and g) and $YZ$-planes (f and h). $XY$-plane is positioned 1~nm above the nanobars, while $YZ$-plane intersects the nanobars at their geometric center. The black contours represent the edges of the a-Si:H bars and the red lines correspond to the substrate surface.}
\label{fig2}
\end{figure*}

The metasurface considered in this work is based on an established design for q-BIC supporting structures \cite{Koshelev:18, Koshelev:19, Loechner:21} consisting of two asymmetric hydrogenated amorphous silicon (a-Si:H) nanobars per unit cell, located on a fused silica substrate. A sketch of the corresponding unit cell is shown in Figs.~${\hyperref[fig2]{\ref{fig2}\text{(a, b)}}}$, respectively. The different widths ($L_{y,1}$ and $L_{y,2}$) of the nanobars allow for the excitation of the q-BIC at the $\Gamma$-point in reciprocal space originating from two in-plane magnetic dipoles oriented in opposite directions and resulting in a non-zero net magnetic dipole moment of the unit cell. The q-BIC-resonance wavelength is designed to coincide with the A$^0$-exciton optical transition of the WSe$_2$ monolayer at room temperature. For metasurface unit cell geometry optimization, we performed numerical simulations based on the finite-element method (FEM), and ellipsometrically measured data for the complex refractive index of a-Si:H. The refractive index of the substrate was taken as 1.45. The optimization resulted in a structure height $h=140$~nm, meta-atom lengths along $x$- and $y$-directions $L_x = 202$~nm, $L_{y,1} = 161$~nm, $L_{y,2} = 140$~nm, separation between meta-atoms $d = 57$~nm, and periodicity along $x$- and $y$-directions $p_x = 367$~nm and $p_y = 413$~nm, respectively.
\par
The validation of fabricated metasurface parameters was performed by scanning-electron microscope (SEM) imaging and result is shown in Fig.~${\hyperref[fig2]{\ref{fig2}\text{(c)}}}$. Next, we conducted a linear-optical transmission spectra measurement of the metasurface using a custom-built setup \cite{Tanaka:20}. These results are shown in Fig.~${\hyperref[fig2]{\ref{fig2}\text{(d)}}}$ and are in good agreement with our theoretical expectations (Fig.~${\hyperref[fig2]{\ref{fig2}\text{(c, d)}}}$). The sharp transmission dip observed for $y$-polarized incident light is associated with the excitation of the q-BIC state. For $x$-polarized incident light, a weaker resonant feature occurs around 700~nm wavelength, well below the spectral range of the excitonic resonance of WSe$_2$ at room temperature centered around 742~nm (red shaded area in Fig.~${\hyperref[fig2]{\ref{fig2}\text{(c, d)}}}$). The near-field intensity profiles at $\lambda = 742$~nm are shown in Fig.~${\hyperref[fig2]{\ref{fig2}\text{(e, g)}}}$ and Fig.~${\hyperref[fig2]{\ref{fig2}\text{(f, h)}}}$ for $x$- and $y$-polarized incident light, respectively. For $y$-polarized incident light, the q-BIC resonance is excited leading to an average electric-field intensity enhancement at 1 nm above the nanobar surface by a factor of 35 with respect to the incident field. In contrast, for $x$-polarized incident light, no q-BIC can be excited at 742 nm, resulting in a smaller average field intensity enhancement of 1.09 within the same plane. As such, we can expect the q-BIC resonance to enhance the interaction of light with a two-dimensional material situated on top of the metasurface.

\subsection{Hybrid Metasurface Assembly and Static Properties}
\begin{figure*}
\centerline{\includegraphics[width = 0.9\linewidth]{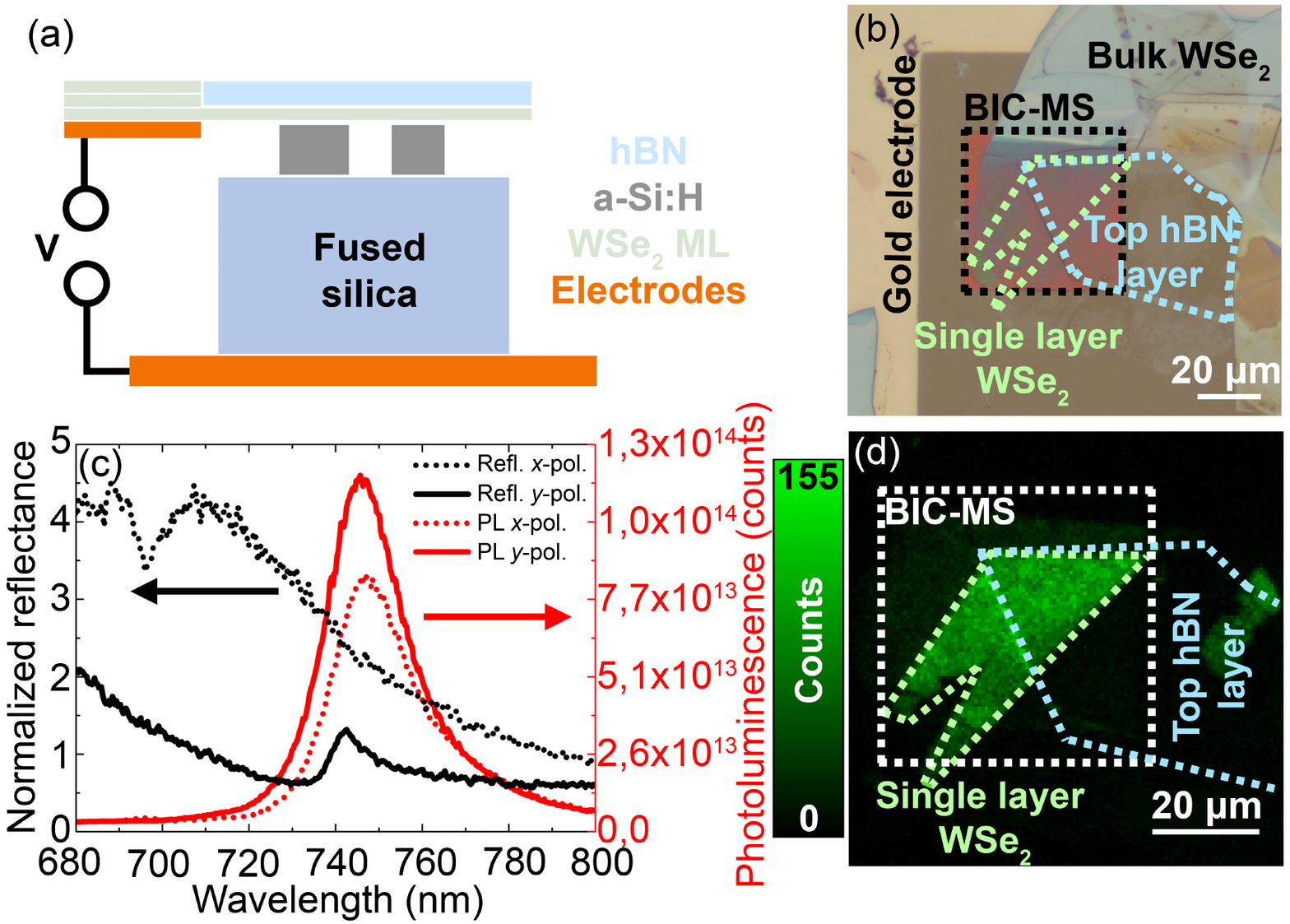}} 
\caption{Hybrid metasurface design and characterization. (a) The schematic of a hybrid metasurface unit cell with contact system represented by top gold injection and bottom copper gate-control electrodes. The top electrode is in direct contact with bulk TMD that is transitioned to a single monolayer covered by few-layer hBN. The potential difference between the electrodes leads to the shift of the Fermi-level in the TMD monolayer (not to scale). (b) Top-view bright-field microscope image of the hybrid resonant structure with main constituents denoted by dash-line borders. The top hBN layer serves for the WSe$_2$ monolayer encapsulation. (c) Linear reflection (black) and PL emission (red) spectra of the hybrid resonant structure for $x$- (dots) and $y$- (solid curves) polarized detected light, respectively. (d) PL emission map of the hybrid resonant structure. The outlines of the WSe$_2$ monolayer, hBN layer, and the metasurface are marked for clarity.}
\label{fig3}
\end{figure*}

The schematic architecture of the hybrid metasurface is shown in Fig.~${\hyperref[fig3]{\ref{fig3}\text{(a)}}}$. The entire sample was placed onto a copper plate serving as a back-gate electrode, while the WSe$_2$ monolayer is in contact with the upper gold electrode through the bulk part of the WSe$_2$ material (depicted as a few layers in Fig.~${\hyperref[fig3]{\ref{fig3}\text{(a)}}}$), and covered by a few-layer hexagonal boron nitride (hBN) crystal for protecting the monolayer part above the active area of the metasurface structure. The potential difference between the WSe$_2$ monolayer and back-gate electrode serves as a driving mechanism of charge doping control within the former. A bright-field microscope image of the hybrid structure is represented in Fig.~${\hyperref[fig3]{\ref{fig3}\text{(b)}}}$ with the main components denoted. Note that the back-gate electrode (not visible in top view) prevents optical measurements in transmission geometry.
\par
To optically characterize the hybrid system without any voltage applied, we measured its linear-optical reflection spectra  for $x$- and $y$-polarized reflected light from the area of the monolayer on top of the metasurface. Corresponding results are dispayed in Fig.~${\hyperref[fig3]{\ref{fig3}\text{(c)}}}$. The expected  correspondence with the transmittance spectra in Fig.~${\hyperref[fig2]{\ref{fig2}\text{(c)}}}$ is clearly observed, with peaks transforming into dips and vice versa. Moreover, we collected polarization-resolved room-temperature PL spectra from the same region. These results are also shown in Fig.~${\hyperref[fig3]{\ref{fig3}\text{(c)}}}$. The small spectral blue shift observed for the PL peak upon changing the linear polarization of excitation from $x$-polarized (748~nm) to $y$-polarized (745~nm) is a first indication of an interaction between the monolayer WSe$_2$ and the metasurface. A corresponding PL map of the system is shown in Fig.~${\hyperref[fig3]{\ref{fig3}\text{(d)}}}$. Outlined are the components of solid-state part of the hybrid system, i.e., the WSe$_2$ monolayer and auxiliary insulating layer of hBN used to separate the monolayer from the atmospheric influence and inhibit its chemical degradation and uncontrolled charge doping. The bright area in the PL map corresponds to the spatial overlap between the metasurface and the monolayer and is caused by the directivity properties of the metasurface, which facilitate the collection of the PL signal.

\subsection{Electrical Gating}
Next, we investigated the behavior of the hybrid system for a variation of the Fermi level of the monolayer subsystem. All experiments were conducted at room temperature.
\par
\begin{figure*}
\centerline{\includegraphics[width = 0.7\linewidth]{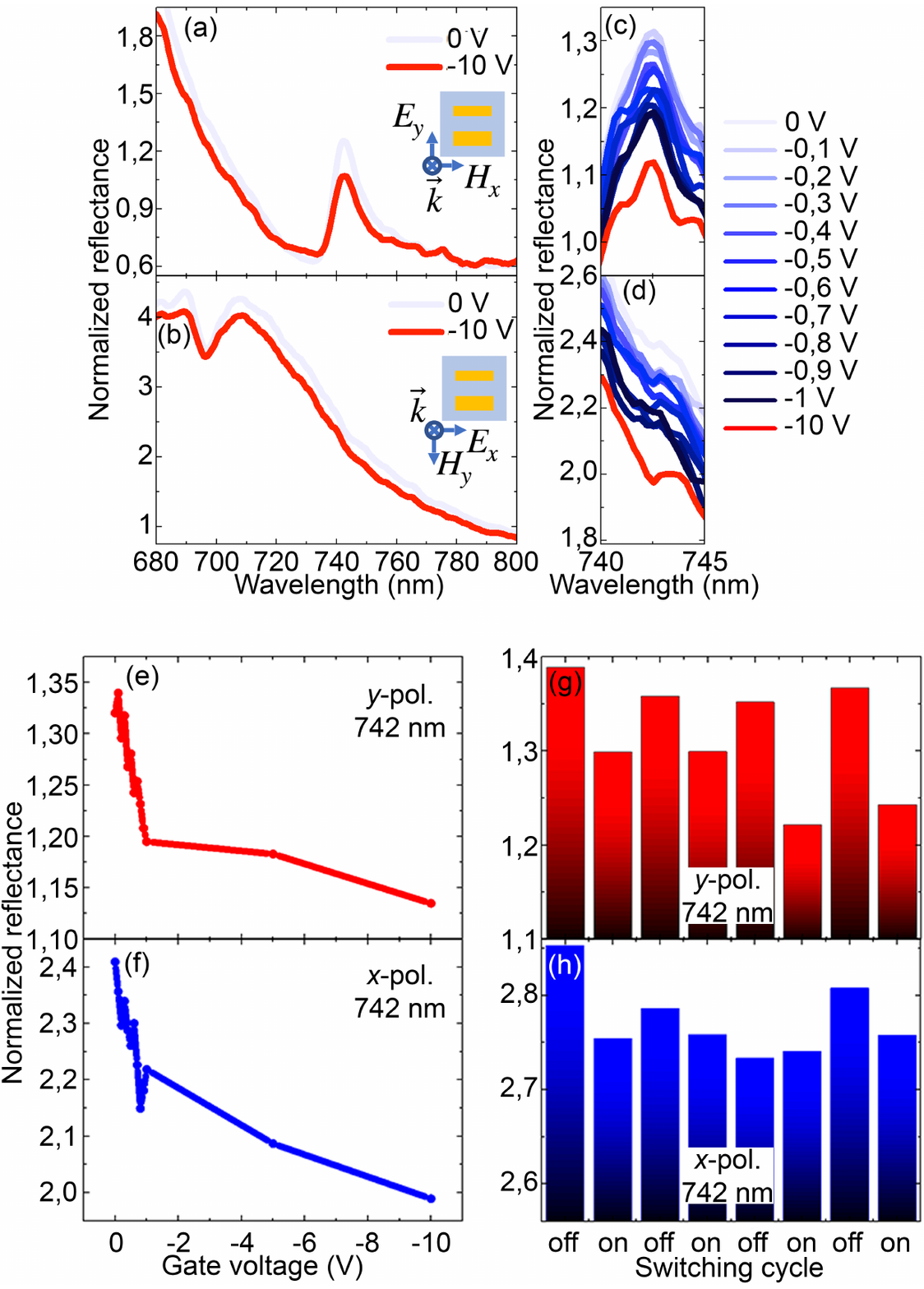}}
\caption{Linear reflectance evolution under external gate voltage. (a, b) Linear reflectance spectra for two polarization states of excitation as depicted in the insets. (c, d) Scaled-up reflectance spectra near A$^{0}$-excitonic transition at 742~nm demonstrating the change of reflectivity for the q-BIC (c) and off-resonant modes (d) when the external gate voltage changes from 0V to -10V. (e, f) The reflectance value at 742 nm for different values of the gate voltage for the q-BIC mode excitation (e) and off-resonant (f) cases. (g, h) The reflectance values at 742 nm for a sequence of gate voltage switching cycles when the A$^{0}$-exciton oscillator strength is suppressed (off-state) and restored (on-state). The reflectance spectra are obtained by normalizing reflection signal from hybrid metasurface to the signal from bare substrate at the same conditions.}
\label{fig4}
\end{figure*}

To this end, we measured its linear-optical reflection spectra using the same setup as described above, but with an external voltage applied between the top injection and bottom gate contacts by an external voltage source (Rohde \& Schwarz NGA 141). Thereby, a negative applied voltage refers to the negative ("-") output of the voltage supply connected to the bottom gate contact and the positive ("+") output connected to the top injection electrode.
These results are shown in Fig.~${\hyperref[fig4]{\ref{fig4}\text{(a, b)}}}$ for $y$- and $x$-polarized reflected signal, respectively, and two different applied voltages that can be considered the switch-on and off states.
\par
The gate voltage restoring the neutral A$^{0}$-exciton absorption characteristics was found as -10 V. While WSe$_2$ monolayers are usually naturally p-doped \cite{Zhang:23} and thus would require a positive gate voltage to reduce the population of the positively charged trion state, this polarity can be explained by inevitable initial n-doping from the substrate and surrounding environment and correspondingly shifted Fermi level for no applied voltage. As such, the "switch-on" state of the neutral exciton is attained for this value and the hybrid structure reaches a global reflectance minimum in the q-BIC resonance. At the zero Fermi level in the system, i.e., when the A$^{0}$-exciton oscillator strength is higher the absorption rate on the corresponding transition increases, leading to a reflection reduction observed near 742~nm (Fig.~${\hyperref[fig4]{\ref{fig4}\text{(a, b)}}}$). Formally, the oscillator strength enhancement of the neutral A$^{0}$-exciton results in an increase in the real part of the 2D conductivity of the WSe$_2$ monolayer, which, in turn, leads to an increase in absorption.

\par
Predicted with our simulations, the modulation depth of reflectance for the case of excited q-BIC state reaches 93\% within the excitonic bandwidth, while for the case when there is no resonant mode excited at the A$^{0}$-excitonic transition it reaches $\approx$ 6\%. This suggests that the reflectance modulation depth is enhanced by the q-BIC resonance at the A$^{0}$-exciton spectral range. In experiments, we observe that both cases can demonstrate the modulation depth of up to 20\%, that can be explained by fabrication imperfections of the metasurface sample and influence of higher-energy oscillators in WSe$_2$ monolayer crystal \cite{Mukherjee:15}. The scaled-up regions of reflectance spectra centered at A$^{0}$-exciton transition for two different excitation schemes (linear polarization of the incident radiation along the $y$- and $x$-axis), and a range of gate voltage values from 0 V to -1 V, with steps of 0.1 V, as well as, the final state at -10 V are represented in Fig.~${\hyperref[fig4]{\ref{fig4}\text{(c, d)}}}$. As it can be observed from the switching dependencies, the evolution of reflectance with gate voltage is strongly influenced by the resonant coupling between the photonic mode of the metasurface and the excitonic transitions.
The reflectance value evolution at 742~nm wavelength as a function of the gate voltage steps is provided in Fig.~${\hyperref[fig4]{\ref{fig4}\text{(e, f)}}}$.
\par
The broad reduction in reflectance across the spectrum, starting from 680~nm, becomes particularly pronounced and begins to compete with excitonic effects at higher gate voltages, starting from 1 V. This behavior, clearly observed in the case of off-resonant excitation in Fig.~${\hyperref[fig4]{\ref{fig4}\text{(b)}}}$, can be attributed to the modulation of the nearest higher-energy oscillator (B-exciton) \cite{Mukherjee:15}, the resonance tail of which inevitably contributes to the modulation of reflectivity.
\par
In order to test the repeatability of the switching mechanism we performed measurements over four switching cycles. Corresponding results for the reflectance value at 742~nm wavelength are shown in Fig.~${\hyperref[fig4]{\ref{fig4}\text{(g, h)}}}$ for $y$- and $x$-polarized reflected light, respectively. The depth of reflectance value modulation is more pronounced for the case of the q-BIC resonance ($y$-polarization) coupled to the A$^{0}$-exciton transition when comparing to the off-resonance case ($y$-polarization). This highlights the importance of coupling between the electronic and photonic subsystems for achieving an enhanced reflectance modulation depth while manipulating the electronic system state in the atomically thin material.

\subsection{Numerical Modeling of the Hybrid Metasurface}
In this section, we discuss the principles underlying the operation of the hybrid metasurface and the impact of charge doping on the 2D-conductivity of the TMD monolayer, which is accompanied by changes in the reflectivity and absorbance properties of the hybrid structure. Our numerical model accounts for the influence of the single-atomic layer of WSe$_2$ positioned on top of the q-BIC metasurface, which was peviously optimized to match the q-BIC resonance with the A$^0$-excitonic resonance of the WSe$_2$ monolayer (see section Metasurface Design and Linear Characterization). To this end, we explicitly model the temperature-dependent Drude-Lorentz model of 2D conductivity in the monolayer TMD that takes into account thermal broadening of excitonic resonances due to intra-band electron-phonon scattering:

\begin{equation}
\sigma_{\text{2D}}(\omega) = - i \varepsilon_{0} \omega h_{\text{eff}} \cdot (\varepsilon_{r}(\omega) - 1),
\label{eq:2d_cond}
\end{equation}

where $\varepsilon_{0}$ is the dielectric permittivity of classical vacuum, $h_{\text{eff}} \approx 6.49~\text{Å}$ corresponds to the effective WSe$_2$ monolayer thickness, and $\varepsilon_{r}$ is its complex dielectric permittivity.

As the TMD monolayer virtually represents a 2D system while being three atoms thick, in accordance with the unit cell geometry of the $D_{3h}^{\sigma}$ point group, the properties of such crystals can be described by effective surface quantities. These quantities can correspond to conventional 3D properties, such as dielectric permittivity, by considering and normalizing them to the effective thickness of a single monolayer. Here, we assume the absence of magneto-optical effects, so the conductivity tensor of the TMD monolayer is diagonal. Given the 2D nature of the material, the out-of-plane components vanish:

\begin{equation}
\sigma_{\text{2D}}=
\begin{bmatrix}
\sigma_{xx} & 0\\
0 & \sigma_{yy}
\end{bmatrix}.
\label{eq:2d_cond_mat}
\end{equation}

The model includes intra- and inter-band contributions of the electronic system to the optical properties of the monolayer. The former term represents metallic behavior of the TMD monolayer in terahertz and optical wavelength ranges, and can be described by the Drude model \cite{You:18}. The semiconductor part that takes into account excitonic contribution to dielectric permittivity is described by the Lorentz model.
\par
Each Lorentz oscillator in the model describes excitonic absorption/reflectance peaks that would be observed experimentally. As mentioned above, when a gate voltage is applied, the Fermi level of the TMD monolayer shifts according to the polarity of the resulting potential difference between the TMD monolayer and the back-gate electrode, as well as the initial doping level of the system. The expression that connects the applied gate voltage with the Fermi level for electrons in the heterostructure is given by \cite{Mak:13}: $E_f = \hbar^2 \pi C V_g / 2 m^* e^3$ with $C$ being the back-gate capacitance, $V_g$ the gate voltage, $m^* \approx 0.36 m_e$ the effective electron mass \cite{Cheiwchanchamnangij:12}, and $e$ the elementary charge. The frequency dependent permittivity of the monolayer material described by the Drude-Lorentz model is given by:

\begin{equation}
\varepsilon_r(\omega)=\varepsilon_{\infty} + \sum_{j} \frac{a_j \omega_P^2}{\omega_j^2 - \omega^2 - i \omega \gamma_j},
\label{eq:dl}
\end{equation}

where $a_j$, $\gamma_j$, and $\omega_j$ are the $j^{\text{th}}$ oscillator's strength, damping factor, and resonant frequency, respectively, $\varepsilon_{\infty} = 1$ corresponds to the static permittivity and $\omega_P = 28.3~\text{meV}$ is the plasma frequency, which depends on the carrier concentration. However, it does not change significantly within the range of electric field values used in this work.
\par
With the known oscillator and static material parameters, the permittivity of a 2D material can be transformed into its optical conductivity using Eq. (\ref{eq:2d_cond}). This transformation enables the material to be treated as a purely 2D sheet while incorporating its optical properties. This approach provides a powerful formalism for modeling excitonic dynamics in external electromagnetic fields and facilitates the optical simulations of hybrid photonic structures \cite{Falkovsky:08}.

The parameters $a_j$, $\gamma_j$, and $\omega_j$ can be obtained either from first-principle calculations or experimentally by performing regression optimization of the model parameters to fit measured absorption, reflection, or transmission spectra. The parameter values acquired from fitting the model to experimental reflectance spectra of the WSe$_2$ monolayer on a fused silica substrate are shown in Table \ref{tab:DL-params}. These parameters include oscillators corresponding to neutral A$^{0}$ ($j = 1$) and charged excitons A$^{\pm}$ ($j = 2$), as well as the higher energy $B$- ($j = 3$) and $C$-states ($j = 4$).

\begin{table}[htbp]
\caption{Parameter values of Drude-Lorentz model for WSe$_2$ monolayer at room temperature}
  \label{tab:DL-params}
  \centering
\begin{tabular}{cccc}
\hline
Oscillator number & $a_j$ & $\omega_j$ (eV) & $\gamma_j$ (eV) \\
\hline
$j = 1$ & $0 - 1.858 \times 10^3$ & 1.67 & 0.046 \\
$j = 2$ & $6.26 \times 10^2$ & 1.653 & 0.068 \\
$j = 3$ & $6.437 \times 10^3$ & 1.919 & 0.098 \\
$j = 4$ & $1.439 \times 10^5$ & 2.485 & 0.313 \\
\hline
\end{tabular}
\end{table}

\par
In addition, we introduce an empirical coefficient taking into account changes in the FWHM and amplitude of excitonic peaks under the influence of an external electric field applied to the heterolayer system. This coefficient also incorporates the temperature of the electronic system and is expressed as $\delta = e^{-12 \pi (E_F - k_b T)^2}$, following the approach described in \cite{Mukherjee:15}. The parameter $\delta$ acts as a multiplier for each oscillator’s damping factor and as a divisor for its strength. The model's initial Fermi level is adjusted in accordance with the experimental observations to represent the specific heterolayer and its initial doping level, which may arise due to fabrication conditions and its surrounding media. In experiments involving applied gate voltage, the initial state of the electronic system influences the balance between neutral (A$^{0}$) and charged (A$^{\pm}$) exciton oscillator strengths, as well as their respective contributions to the dielectric permittivity of the TMD monolayer.
\par
\begin{figure*}
\centerline{\includegraphics[width = 0.9\linewidth]{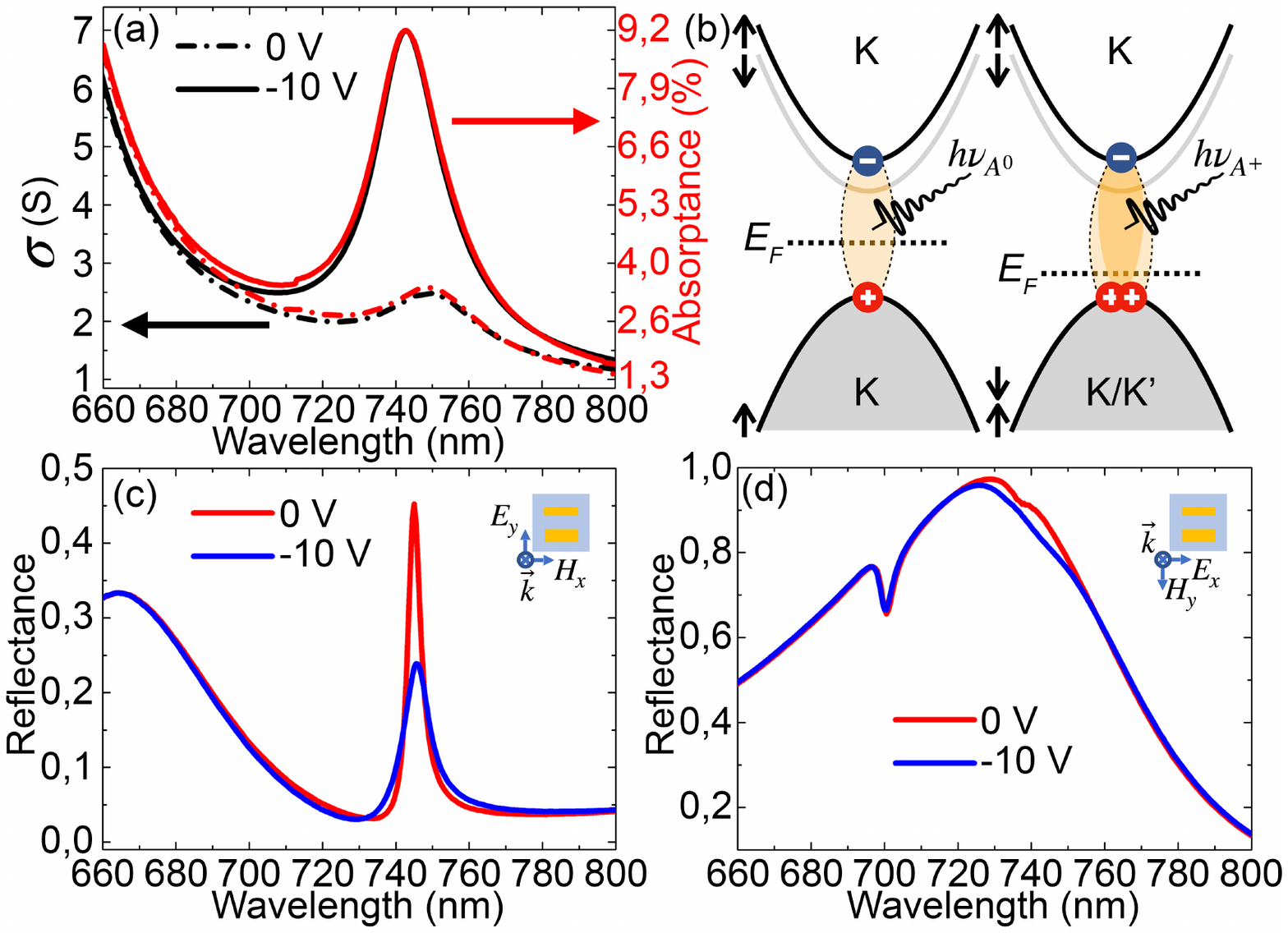}}
\caption{Calculated optical properties of a WSe$_2$ monolayer and the hybrid metasurface. (a) Two-dimensional conductivity values, calculated based on the 2D-conductance model for two gate voltage values (black), along with the corresponding absorptance spectra of the WSe$_2$ monolayer atop the fused silica substrate (red). (b) Illustration of Fermi-level control mechanism of the $A^{0}$-exciton and $A^{\pm}$-trion balance in optical response of WSe$_2$ monolayer. It demonstrates the formation of the positively charged $A^+$-trion when the Fermi level shifts towards the upper valence band. (c, d) Calculated reflectance spectra of the hybrid metasurface upon two different gate voltage values for the q-BIC (c) and off-resonant (d) cases. The excitation polarization schemes are illustrated in the insets.}
\label{fig5}
\end{figure*}

The real part of the sheet conductivity of optimized numerical model is shown in Fig.~${\hyperref[fig5]{\ref{fig5}\text{(a)}}}$.
Taking into account the initial doping level of the WSe$_2$ monolayer, switching the gate voltage from 0 V to -10 V restores the A$^{0}$-exciton contribution to the 2D conductivity. This restoration leads to an effective blue shift of the overall excitonic peak and an increase in its amplitude, resulting in higher absorption, as shown in Fig.~${\hyperref[fig5]{\ref{fig5}\text{(a)}}}$. The schematic band structure and two different states of electronic system of the WSe$_2$ monolayer (A$^0$-exciton and A$^+$-trion) are represented in Fig.~${\hyperref[fig5]{\ref{fig5}\text{(b)}}}$. When the Fermi level is positioned in the middle of the band gap, it corresponds to a restored population of neutral A$^0$-excitons, as the charge densities of electrons and holes become equalized. When the negative gate potential is further applied and the Fermi level shifts towards the upper valence band, the relative concentration of positively charged holes effectively increases, leading to a higher probability of forming the positively charged trionic state  A$^{+}$. The bright $A^{+}$ states are formed by one electron near the $K$($K'$)-point and two holes from different valleys $K$/$K'$ ($K'$/$K$) \cite{Drueppel:17}. Such a configuration results in a lower binding energy compared to the $A^0$ state, which, in its turn, entails a lower oscillator strength and a reduced radiative decay rate, when described by the Drude-Lorentz model. As a result, trionic states in monolayer TMDs manifest in their optical properties as broader and weaker resonances compared to neutral excitons \cite{Mak:13}. The opposite gate potential facilitates the development of negatively charged trions $A^-$, which have a binding energy nearly identical to that of $A^+$ states (within $\sim 1~\text{meV}$ difference). As a result, $A^-$ trions exhibit effectively the same total energy and contribute similarly to the 2D dielectric function of the WSe$_2$ monolayer. This implies that, when normalized to the initial shift of the Fermi level in the WSe$_2$ monolayer crystal, the optical properties of the hybrid metasurface would exhibit similar behavior in both negative and positive gate potential ranges at room temperature.
\par
The increase in oscillator strength of the neutral A$^{0}$-exciton leads to higher absorption in WSe$_2$ monolayer and, as a result, to the lower reflection coefficient of the hybrid metasurface within the spectral range of the excitonic transition (Fig.~${\hyperref[fig5]{\ref{fig5}\text{(c, d)}}}$). Although the effect is observed for both the q-BIC mode and off-resonance cases, the influence of the q-BIC mode is particularly pronounced, demonstrating nearly double the modulation depth of the switching effect, reaching $\approx 93~\%$ for excitonic transitions coupled to the q-BIC state. A stronger modulation effect could be achieved with a higher $Q$-factor resonance and at lower temperatures that would reduce the thermal broadening of excitonic peaks.
\par
The calculation model, based on the described considerations, demonstrates a good qualitative agreement with experimental results for the hybrid metasurface, as evident from measured and calculated reflectance spectra shown in Fig.~${\hyperref[fig4]{\ref{fig4}\text{(a, b)}}}$ and Fig.~${\hyperref[fig5]{\ref{fig5}\text{(c, d)}}}$, correspondingly. It quantitatively describes excitonic tuning of a WSe$_2$ monolayer and its influence on the reflectance change of the hybrid metasurface upon exertion of an external electric field.

\section{Discussion}
We have experimentally demonstrated voltage tuning of a resonant metasurface empowered by a TMD monolayer. To this end, we have hybridized a dielectric metasurface supporting a q-BIC resonance at the neutral A$^{0}$-exciton wavelength of WSe$_2$ with a monolayer crystal of this material. Application of a voltage allowed us to reversibly change the Fermi level in the monolayer crystal. In this way, we induced a change of the population of its excitonic and trionic states and thereby to a variation of the frequency-dependent damping factor and our experimental observable, the reflectance of the hybrid metasurface. The effect is enhanced by coupling between the WSe$_2$ and the q-BIC resonances. While it still remains small, reaching only $\approx 20~\%$ modulation of the initial transmission level of the q-BIC resonance, it may be enhanced in the future by improved designs and integration strategies, further strengthening the coupling between the photonic mode and the excitonic resonance. Our experimental results are underpinned by numerial simulations, which quantitatively describe the excitonic dynamics in the WSe$_2$ monolayer and its influence on the linear reflectance change of the hybrid metasurface for an applied electric field. A good overall agreement is obtained between experimental and numerical results, positioning our model as a powerful tool for describing and prototyping tunable resonant photonic structures based on Fermi level changes in 2D-TMDs.
\par
Altogether, our work introduces a new class of actively tunable
photonic nanostructures that may find practical applications as adaptive optical elements, e.g. in holographic systems or optical neural networks.

\section{Materials and methods}

\subsection{Numerical simulations}
All numerical simulations in this work were made using the software package COMSOL Multiphysics 6.0 (Wave Optics module), which is based on the finite-element method (FEM) \cite{comsol}.

\subsection{Linear optical characterization}
For measuring linear-optical reflection spectra we used a stabilized free-space light source (Thorlabs SLS303) integrated with reflection-geometry microscope based on ZEISS Axio Observer. The sample orientation was controlled by moving the stage of the microscope. To spatially filter the light reflected from the sample, an adjustable rectangular-shaped iris was placed in the focal plane of the objective, which collected the reflected signal from the microscope system. The quality of the filtered signal was controlled by a CCD-camera. After filtering, the signal was coupled into a multimode fiber connected to a spectrometer (ANDOR Kymera 328i-D2-sil). The measured spectra were normalized to the reflection from a clean region of the bare substrate (fused silica).
\par
PL spectra were aquired using a commercially available PicoQuant MicroTime 200 confocal microscopy platform in combination with the same spectrometer.

\subsection{Nanofabrication}
The fabrication of silicon-based q-BIC structure started with the deposition of a 135-nm-thick a-Si:H film on a silica substrate. Plasma-enhanced chemical vapor deposition (PECVD) was carried out for the film in Plasmalab 100 from Oxford in which silane (SiH$_4$) precusor and helium (He) dilution gas were used. To define the desired features of metasurface, a positive electron beam resist (ZEP520A from Zeon Chemicals) was spin-coated onto the sample. In order to prevent electron charging during the electron-beam exposure process, a coating of Espacer (300Z from Showa Denko) was applied on the resist. The nano-bar patterns were written using Raith150 system, followed by the development of the resist in ZED-N50. Then, a 30-nm-thick aluminum film was deposited on the sample through electron-beam evaporation (Temescal BJD-2000), and was patterned by removing the resist with resist remover (ZDMAC from Zeon Co.). Obtained aluminum patterns served as a hard etch mask during plasma etching of a-Si:H. The etch process was carried out using a fluorine-based inductively coupled-plasma reactive ion etching system (Oxford Plasmalab System 100), where the etching conditions were optimized by precisely controlling the mixing ratio of CHF$_3$ and SF$_6$ gases for vertical etch profile. Subsequently, aluminum wet etchant was applied to remove any residual aluminum from the patterned a-Si:H structure.
\par
The electrodes for WSe$_2$ monolayer doping level control were fabricated by means of DMD-based maskless lithography system (Smart Print UV) that was used for printing a positive mask (AZ1518 positive photoresist) for physical vapor deposition (PVD) of a 30-nm thick gold layer. The printing resolution was 2 $\mu$m. The lift-off process of remaining metal layer was conducted in acetone solution and assisted by few-hours ultrasonication.
\par
Following the procedure similar to \cite{Castellanos:14}, hybridization of the metasurface with the 2D materials was performed by tape exfoliation method of WSe$_2$ and hBN bulk crystals. The thickness of the exfoliated layers was controlled by contrast measurements \cite{Mueller:15} using an optical microscope in visible range, and monolayer quality was double-checked with PL-measurements. The exfoliated crystals were collected layer by layer with a combined polydimethylsiloxane (PDMS)-polycarbonate (PC) stamp that ensures clean and robust adhesion for pick-up and deposition of collected multi-layered systems onto various substrate materials and nanostructures. The deposition of the multi-layered structure onto the metasurface was carried out at constant substrate temperature of 150 $^{\circ}$C and kept at this elevated value for several hours to allow the PDMS-PC stamp to stabilize and for the PC layer to detach from the PDMS part. After deposition with the PC film, the multi-layered structure was cleaned in trichloromethane (chloroform) solution for several hours until the PC layer is completely dissolved.

\section{Acknowledgments}
This work was funded by the Deutsche Forschungsgemeinschaft (DFG, German Research Foundataion) - CRC/SFB 1375 NOA ”Nonlinear Optics down to Atomic scales” (project number 398816777), International Research Training Group 2675 “META-Active” (project number 437527638), and project number 448835038. A.B. gratefully acknowledges financial support from Spanish national project No. PID2022-137857NA-I00. A.B. thanks MICINN for the Ramon y Cajal Fellowship (grant No. RYC2021-030880-I).

\section{Data availability statement}
The data is available from authors upon reasonable request.

\section{Conflict of interest}
The authors declare no competing interests.


\begin{thebibliography}{99}

\bibitem{Elshaari:20} Elshaari, A. W., Pernice, W., Srinivasan, K., Benson, O. \& Zwiller, V. Hybrid integrated quantum photonic circuits. \href{https://doi.org/10.1038/s41566-020-0609-x}{\textit{Nat. Photonics}} \textbf{14}, 285--298 (2020).

\bibitem{Bogaerts:20} Bogaerts, W. et al. Programmable photonic circuits. \href{https://doi.org/10.1038/s41586-020-2764-0}{\textit{Nature}} \textbf{586}, 207--216 (2020).

\bibitem{Wang:15} Wang, G. et al. Giant Enhancement of the Optical Second-Harmonic Emission of ${\mathrm{WSe}}_{2}$ Monolayers by Laser Excitation at Exciton Resonances. \href{https://link.aps.org/doi/10.1103/PhysRevLett.114.097403}{\textit{Phys. Rev. Lett.}} \textbf{114}, 097403 (2015).

\bibitem{Cheiwchanchamnangij:12} Cheiwchanchamnangij, T. \& Lambrecht, W. R. L. Quasiparticle band structure calculation of monolayer, bilayer, and bulk MoS${}_{2}$. \href{https://link.aps.org/doi/10.1103/PhysRevB.85.205302}{\textit{Phys. Rev. B}} \textbf{85}, 205302 (2012).

\bibitem{Chernikov:14} Chernikov, A. et al. Exciton Binding Energy and Nonhydrogenic Rydberg Series in Monolayer ${\mathrm{WS}}_{2}$. \href{https://link.aps.org/doi/10.1103/PhysRevLett.113.076802}{\textit{Phys. Rev. Lett.}} \textbf{113}, 076802 (2014).

\bibitem{Wang:12} Wang, Q. H., Kalantar-Zadeh, K., Kis, A., Coleman, J. N. \& Strano, M. S. Electronics and optoelectronics of two-dimensional transition metal dichalcogenides. \href{https://doi.org/10.1038/nnano.2012.193}{\textit{Nature Nanotech.}} \textbf{7}, 699--712 (2012).

\bibitem{Klimmer:21} Klimmer, S. et al. All-optical polarization and amplitude modulation of second-harmonic generation in atomically thin semiconductors. \href{https://doi.org/10.1038/s41566-021-00859-y}{\textit{Nat. Photonics}} \textbf{15}, 837--842 (2021).

\bibitem{Sinelnik:24} Sinelnik, A. et al. Ultrafast all-optical second harmonic wavefront shaping. \href{https://doi.org/10.1038/s41467-024-46642-9}{\textit{Nat. Commun.}} \textbf{15}, 2507 (2024).

\bibitem{Ryder:16} Ryder, C. R., Wood, J. D., Wells, S. A. \& Hersam, M. C. Chemically Tailoring Semiconducting Two-Dimensional Transition Metal Dichalcogenides and Black Phosphorus. \href{https://doi.org/10.1021/acsnano.6b01091}{\textit{ACS Nano}} \textbf{10,} 3900--3917 (2016).

\bibitem{Seyler:15} Seyler, K. L. et al. Electrical control of second-harmonic generation in a WSe2 monolayer transistor. \href{https://doi.org/10.1038/nnano.2015.73}{\textit{Nature Nanotech.}} \textbf{15}, 407--411 (2015).

\bibitem{Yu:17} Yu, Y. et al. Giant Gating Tunability of Optical Refractive Index in Transition Metal Dichalcogenide Monolayers. \href{https://doi.org/10.1021/acs.nanolett.7b00768}{\textit{Nano Lett.}} \textbf{17}, 3613--3618 (2017).

\bibitem{Chernikov:15} Chernikov, A. et al. Electrical Tuning of Exciton Binding Energies in Monolayer ${\mathrm{WS}}_{2}$. \href{https://link.aps.org/doi/10.1103/PhysRevLett.115.126802}{\textit{Phys. Rev. Lett.}} \textbf{115}, 126802 (2015).

\bibitem{Cunningham:17} Cunningham, P. D., Hanbicki, A. T., McCreary, K. M. \& Jonker, B. T. Photoinduced Bandgap Renormalization and Exciton Binding Energy Reduction in WS$_2$. \href{https://doi.org/10.1021/acsnano.7b06885}{\textit{ACS Nano}} \textbf{11}, 12601--12608 (2017).

\bibitem{Fan:17} Fan, X. et al. Nonlinear photoluminescence in monolayer WS2: parabolic emission and excitation fluence-dependent recombination dynamics. \href{http://dx.doi.org/10.1039/C7NR01345K}{\textit{Nanoscale}} \textbf{9}, 7235--7241 (2017).

\bibitem{Peng:20} Peng, Z., Chen, X., Fan, Y., Srolovitz, D. J. \& Lei, D. Strain engineering of 2D semiconductors and graphene: from strain fields to band-structure tuning and photonic applications. \href{https://doi.org/10.1038/s41377-020-00421-5}{\textit{Light Sci. Appl.}} \textbf{9}, 190 (2020).

\bibitem{Raja:17} Raja, A. et al. Coulomb engineering of the bandgap and excitons in two-dimensional materials. \href{https://doi.org/10.1038/ncomms15251}{\textit{Nat. Commun.}} \textbf{8}, 15251 (2017).

\bibitem{Stier:16} Stier, A. V., Wilson, N. P., Clark, G., Xu, X. \& Crooker, S. A. Probing the Influence of Dielectric Environment on Excitons in Monolayer WSe2: Insight from High Magnetic Fields. \href{https://doi.org/10.1021/acs.nanolett.6b03276}{\textit{Nano Lett.}} \textbf{16}, 7054--7060 (2016).

\bibitem{Kobayashi:15} Kobayashi, Y. et al. Growth and Optical Properties of High-Quality Monolayer WS2 on Graphite. \href{https://doi.org/10.1021/acsnano.5b00103}{\textit{ACS Nano}} \textbf{9}, 4056--4063 (2015).

\bibitem{Lanzillo:13} Lanzillo, N. A. et al. Temperature-dependent phonon shifts in monolayer MoS$_2$. \href{https://doi.org/10.1063/1.4819337}{\textit{Appl. Phys. Lett.}} \textbf{103}, 093102 (2013).

\bibitem{Thripuranthaka:14} Thripuranthaka, M. \& Dattatray J. L. Temperature Dependent Phonon Shifts in Single-Layer WS$_2$. \href{https://doi.org/10.1021/am404847d}{\textit{ACS Appl. Mater. Interfaces}} \textbf{6}, 1158--1163 (2014).

\bibitem{Li:21} Li, M., Biswas, S., Hail, C. U. \& Atwater, H. A. Refractive Index Modulation in Monolayer Molybdenum Diselenide. \href{https://doi.org/10.1021/acs.nanolett.1c02199}{\textit{Nano Lett.}} \textbf{21}, 7602--7608 (2021).

\bibitem{Chen:17} Chen, H. et al. Enhanced second-harmonic generation from two-dimensional MoSe$_2$ on a silicon waveguide. \href{https://doi.org/10.1038/lsa.2017.60}{\textit{Light Sci. Appl.}} \textbf{6}, e17060 (2017).

\bibitem{Chen:20} Chen, Y. et al. Metasurface Integrated Monolayer Exciton Polariton. \href{https://doi.org/10.1021/acs.nanolett.0c01624}{\textit{Nano Lett.}} \textbf{20}, 5292--5300 (2020).

\bibitem{Loechner:21} Löchner, F. J. F. et al. Hybrid Dielectric Metasurfaces for Enhancing Second-Harmonic Generation in Chemical Vapor Deposition Grown MoS2 Monolayers. \href{https://doi.org/10.1021/acsphotonics.0c01375}{\textit{ACS Photonics}} \textbf{8}, 218--227 (2021).

\bibitem{Hsu:13} Hsu, C. W. et al. Observation of trapped light within the radiation continuum. \href{https://doi.org/10.1038/nature12289}{\textit{Nature}} \textbf{499}, 188--191 (2013).

\bibitem{Kodigala:17} Kodigala, A. et al. Lasing action from photonic bound states in continuum. \href{https://doi.org/10.1038/nature20799}{\textit{Nature}} \textbf{541}, 196--199 (2017).

\bibitem{Bernhardt:20} Bernhardt, N. et al. Quasi-BIC resonant enhancement of second-harmonic generation in WS$_2$ monolayers. \href{https://doi.org/10.1021/acs.nanolett.0c01603}{\textit{Nano Lett.}} \textbf{20}, 5309--5314 (2020).

\bibitem{Kim:22_3} Kim, Y., Three-dimensional artificial chirality towards low-cost and ultra-sensitive enantioselective sensing. \href{http://dx.doi.org/10.1039/D1NR05805C}{\textit{Nanoscale}} \textbf{14}, 3720--3730 (2022).

\bibitem{Gorkunov:20} Gorkunov, M. V., Antonov, A. A. \& Kivshar, Y. S. Metasurfaces with maximum chirality empowered by bound states in the continuum. \href{https://link.aps.org/doi/10.1103/PhysRevLett.125.093903}{\textit{Phys. Rev. Lett.}} \textbf{125}, 093903 (2020).

\bibitem{Zhu:20} Zhu, L., Yuan, S., Zeng, C. \& Xia, J. Manipulating photoluminescence of carbon G–center in silicon metasurface with optical bound states in the continuum. \href{https://doi.org/10.1002/adom.201901830}{\textit{Adv. Opt. Mater.}} \textbf{8}, 1901830 (2020).

\bibitem{Hsu:19} Hsu, C. W., Zhen, B., Stone, A. D., Joannopoulos, J. D. \& Soljačić, M. M. Bound states in the continuum. \href{https://doi.org/10.1038/natrevmats.2016.48}{\textit{Nat. Rev. Mater.}} \textbf{1}, 16048 (2016).

\bibitem{Koshelev:19} Koshelev, K., Bogdanov, A. \& Kivshar, Y. Meta-optics and bound states in the continuum. \href{https://doi.org/10.1016/j.scib.2018.12.003}{\textit{Sci. Bull.}} \textbf{64}, 836--842 (2019).

\bibitem{Han:19} Han, S. et al. All-Dielectric Active Terahertz Photonics Driven by Bound States in the Continuum. \href{https://doi.org/10.1002/adma.201901921}{\textit{Adv. Mater.}} \textbf{31}, 1901921 (2019).

\bibitem{Abujetas:19} Abujetas, D. R. et al. Brewster quasi bound states in the continuum in all-dielectric metasurfaces from single magnetic-dipole resonance meta-atoms. \href{https://doi.org/10.1002/adma.201901921}{\textit{Sci. Rep.}} \textbf{9}, 16048 (2019).

\bibitem{Wu:15} Wu, S. et al. Monolayer semiconductor nanocavity lasers with ultralow thresholds. \href{https://doi.org/10.1038/nature14290}{\textit{Nature}} \textbf{520}, 69--72 (2015).

\bibitem{Li:17} Li, Y. et al. Room-temperature continuous-wave lasing from monolayer molybdenum ditelluride integrated with a silicon nanobeam cavity. \href{https://doi.org/10.1038/nnano.2017.128}{\textit{Nature Nanotech.}} \textbf{12}, 987--992 (2017).

\bibitem{Li:23} Li, M., Hail, C. U., Biswas, S. \& Atwater, H. A. Excitonic Beam Steering in an Active van der Waals Metasurface. \href{https://doi.org/10.1021/acs.nanolett.3c00032}{\textit{Nano Letters}} \textbf{23}, 2771--2777 (2023).

\bibitem{Zhao:20} Zhao, P. et al. Electronic and optical properties of transition metal dichalcogenides under symmetric and asymmetric field-effect doping. \href{https://dx.doi.org/10.1088/1367-2630/aba8d2}{\textit{New Journal of Physics}} \textbf{22}, 083072 (2020).

\bibitem{Mukherjee:15} Mukherjee, B. et al. Complex electrical permittivity of the monolayer molybdenum disulfide (MoS$_2$) in near UV and visible. \href{https://doi.org/10.1364/OME.5.000447}{\textit{Opt. Mater. Express}} \textbf{5}, 447--455 (2015).

\bibitem{Koshelev:18} Koshelev, K., Lepeshov, S., Liu, M., Bogdanov, A. \& Kivshar, Y. Asymmetric Metasurfaces with High-$Q$ Resonances Governed by Bound States in the Continuum. \href{https://link.aps.org/doi/10.1103/PhysRevLett.121.193903}{\textit{Phys. Rev. Lett.}} \textbf{121}, 193903 (2018).

\bibitem{Tanaka:20} Tanaka, K. et al. Chiral Bilayer All-Dielectric Metasurfaces. \href{https://doi.org/10.1021/acsnano.0c07295}{\textit{ACS Nano}} \textbf{14}, 15926--15935 (2020).

\bibitem{Zhang:23} Zhang, Y.-Z., Zhu, G.-J. \& Yang, J.-H. Origin of p-type conductivity in a WSe2 monolayer. \href{http://dx.doi.org/10.1039/D3NR01321A}{\textit{Nanoscale}} \textbf{15}, 12116--12122 (2023).

\bibitem{You:18} You, J. W., Threlfall, E., Gallagher, D. F. G. \& Panoiu, N. C. Computational analysis of dispersive and nonlinear 2D materials by using a GS-FDTD method.  \href{https://doi.org/10.1364/JOSAB.35.002754}{\textit{J. Opt. Soc. Am. B}} \textbf{35}, 2754--2763 (2018).

\bibitem{Mak:13} Mak, K. F. et al. Tightly bound trions in monolayer MoS$_2$. \href{https://doi.org/10.1038/nmat3505}{\textit{Nature Mater.}} \textbf{12}, 207--211 (2013).

\bibitem{Falkovsky:08} Falkovsky L.A. Optical properties of graphene. \href{http://dx.doi.org/10.1088/1742-6596/129/1/012004}{\textit{J. Phys.: Conf. Ser.}} \textbf{129}, 012004 (2008).

\bibitem{Drueppel:17} Drüppel, M., Deilmann, T., Krüger, P., \& Rohlfing, M. Diversity of trion states and substrate effects in the optical properties of an MoS$_2$ monolayer \href{https://doi.org/10.1038/s41467-017-02286-6}{\textit{Nat. Commun.}} \textbf{8}, 2117 (2017).

\bibitem{comsol} The official COMSOL Multiphysics\textsuperscript\textregistered web-site \href{https://www.comsol.com}{https://www.comsol.com}.

\bibitem{Castellanos:14} Castellanos-Gomez, A. et al. Deterministic transfer of two-dimensional materials by all-dry viscoelastic stamping. \href{http://dx.doi.org/10.1088/2053-1583/1/1/011002}{\textit{2D Mater.}} \textbf{1}, 011002 (2014).

\bibitem{Mueller:15} Müller, M. R. et al. Visibility of two-dimensional layered materials on various substrates. \href{https://doi.org/10.1063/1.4930574}{\textit{J. Appl. Phys.}} \textbf{118}, 145305 (2015).

\end{thebibliography}
\end{document}